\newcommand{\sysname}{\textsc{raggy}\xspace}

\documentclass[sigconf, nonacm]{acmart}

\usepackage{xspace} 
\usepackage{rotating}
\usepackage{csquotes}
\usepackage{multirow}

\renewcommand{\blockquote}[1]{``\textit{#1}''}

\usepackage{pgfplots}
\pgfplotsset{compat=1.18}
\usepackage{pgfplotstable}
\usepackage{xcolor}

\usepackage{enumitem}
\usepackage{cleveref}

\usepackage{array}

\usepackage{listings}
\usepackage{xcolor}

\definecolor{darkgreen}{RGB}{0,100,0}  
\definecolor{paramcolor}{RGB}{128,0,128}  
\definecolor{bluecolor}{RGB}{0,0,255}  
\definecolor{redcolor}{RGB}{200,0,0}   
\definecolor{yellowcolor}{RGB}{180,140,0} 
\definecolor{numbercolor}{RGB}{0,128,0} 
\definecolor{graycolor}{RGB}{150,150,150}
\definecolor{lightgray}{RGB}{245, 245, 245}  %

\lstdefinestyle{python}{
    backgroundcolor=\color{lightgray},  
    commentstyle=\color{graycolor},      
    keywordstyle=\bfseries\color{darkgreen},  %
    stringstyle=\color{redcolor},  %
    identifierstyle=\color{black},  
    emph={retriever, llm, Query, Answer}, emphstyle=\color{bluecolor}, %
    emph={[2]invoke, format}, emphstyle={[2]\color{yellowcolor}}, %
    basicstyle=\ttfamily\small,     
    breakatwhitespace=false,
    breaklines=true,
    captionpos=b,
    keepspaces=true,
    numbers=left, numberstyle=\tiny\color{gray},  %
    showspaces=false,
    showstringspaces=false,
    showtabs=false,
    tabsize=2,
    morekeywords={k, chunkSize, chunkOverlap, searchBy, max_tokens, temperature, retrievalMode}, 
    morekeywords=[2]{0, 1, 2, 3, 4, 5, 6, 7, 8, 9},  %
    keywordstyle=[2]\color{numbercolor}, %
}

\AtBeginDocument{%
  }

\setcopyright{acmlicensed}
\copyrightyear{2018}
\acmYear{2018}
\acmDOI{XXXXXXX.XXXXXXX}

\acmConference[Conference acronym 'XX]{Make sure to enter the correct
  conference title from your rights confirmation emai}{June 03--05,
  2018}{Woodstock, NY}

\begin{document}

\title{RAG Without the Lag: Interactive Debugging for Retrieval-Augmented Generation Pipelines}

\author{Quentin Romero Lauro\textsuperscript{*}$^1$, Shreya Shankar\textsuperscript{*}$^2$, Sepanta Zeighami$^2$, Aditya G. Parameswaran$^2$}
\affiliation{%
$^1$University of Pittsburgh, $^2$UC Berkeley \\
\url{quentinrl @ pitt.edu}, \{\url{shreyashankar, zeighami, adityagp}\} \url{@ berkeley.edu}\country{}}

\newif\ifanonymous
\anonymoustrue %
\ifanonymous
\else
  \thanks{$^\ast$Co-first author contribution. This work was conducted while Q. Romero Lauro was participating in the Research Experience for Undergraduates (REU) program at UC Berkeley EECS.}
\fi

\renewcommand{\shortauthors}{Romero Lauro and Shankar et al.}
\newcommand{\shreya}[1]{\textcolor{teal}{[Shreya: #1]}}
\newcommand{\quentin}[1]{\textcolor{blue}{[Quentin: #1]}}
\newcommand{\agp}[1]{\textcolor{red}{[Aditya: #1]}}
\newcommand{\sep}[1]{\textcolor{orange}{[Sep: #1]}}

\newcommand{\topic}[1]{\vspace{-3.5pt}\smallskip \smallskip \noindent{\bf #1.}}
\newcommand{\ttt}[1]{{\small \texttt{#1}}\xspace}

\newcommand{\papertext}[1]{#1}
\newcommand{\techreport}[1]{}

\begin{abstract}
Retrieval-augmented generation (RAG) pipelines have become the de-facto approach for building AI assistants with access to external, domain-specific knowledge. Given a user query, RAG pipelines typically first retrieve (\underline{R}) relevant information from external sources, before invoking a Large Language Model (LLM), augmented (\underline{A}) with this information, to generate (\underline{G}) responses. Modern RAG pipelines frequently chain multiple retrieval and generation components, in any order. However, developing effective RAG pipelines is challenging because retrieval and generation components are intertwined, making it hard to identify which component(s) cause errors in the eventual output. The parameters with the greatest impact on output quality often require hours of pre-processing after each change, creating prohibitively slow feedback cycles. To address these challenges, we present \sysname, a developer tool that combines a Python library of composable RAG primitives with an interactive interface for real-time debugging. We contribute the design and implementation of \sysname, insights into expert debugging patterns through a qualitative study with 12 engineers, and design implications for future RAG tools that better align with developers' natural workflows.
\end{abstract}



\begin{teaserfigure}
\centering
\includegraphics[width=0.9\linewidth]{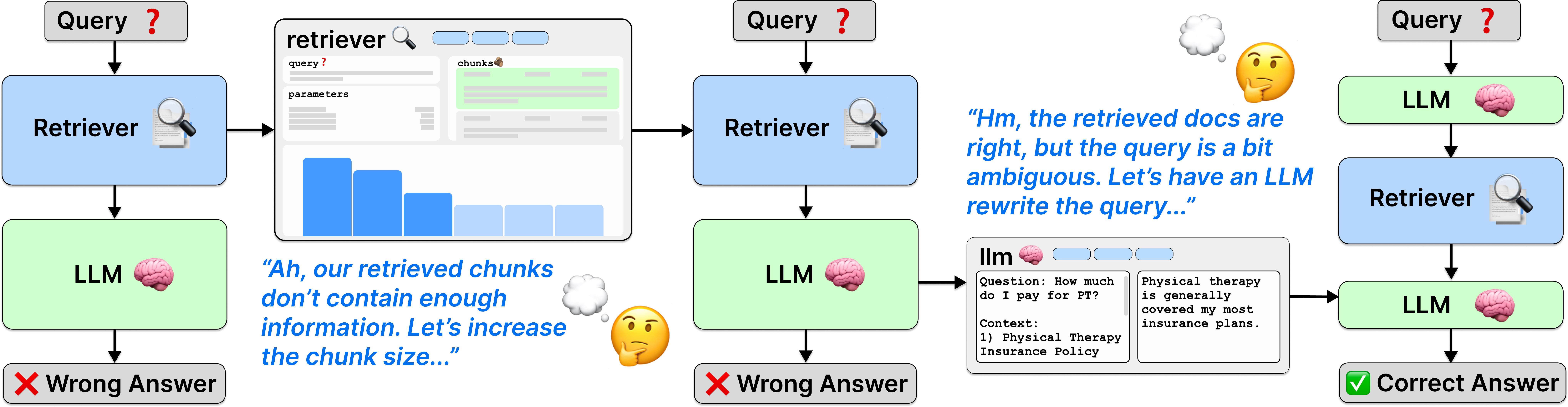}
\Description[A diagram showing the iterative debugging process of a RAG pipeline using RAGGY]{The image shows a three-panel diagram illustrating how users can debug and improve a RAG (Retrieval Augmented Generation) pipeline using RAGGY. Each panel represents a pipeline configuration. The left panel shows a pipeline with 3 steps: basic query, retriever, then LLM  resulting in a ``Wrong Answer'' with a user thought bubble indicating insufficient information in retrieved chunks. The middle panel shows a modified pipeline after increasing chunk size, but still producing a wrong answer with a speech bubble noting ``Hm, the retrieved docs are right, but the query is a bit ambiguous. Let's have an LLM rewrite the query...'' The right panel shows a successful pipeline with query, retriever, LLM, retriever, and LLM components resulting in a Correct Answer. This visualization demonstrates how RAGGY enables real-time experimentation with immediate feedback to optimize RAG pipelines.}
\caption{An example of how a user might use \sysname, our tool for debugging RAG pipelines. The developer debugs by first increasing chunk size to retrieve more information, then by adding a new LLM step to rewrite the query to address ambiguity. \sysname enables real-time modifications with instant feedback, and facilitates visualization of how changes impact the pipeline.}
\label{fig:iterationwithraggy}
\end{teaserfigure}

\maketitle

\begin{figure*}
\centering
\includegraphics[width=0.8\linewidth]{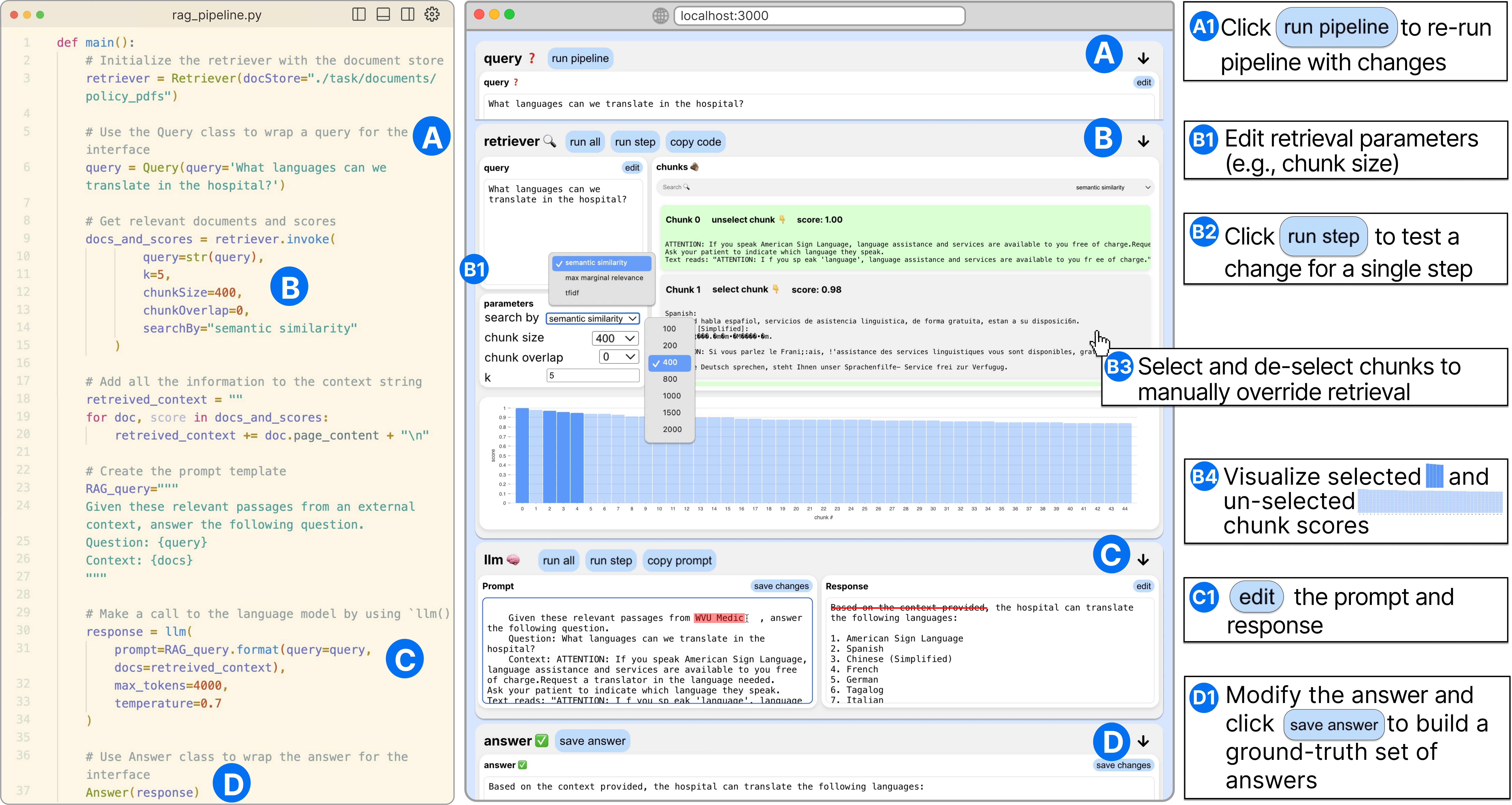}
\Description[Side-by-side comparison of Python RAG implementation code and RAGGY debugging interface]{The figure shows a side-by-side comparison of Python code for a RAG pipeline (left) and RAGGY's interactive debugging interface (right). The left panel displays Python code with color-highlighted sections labeled A-D that implement different components of the RAG pipeline. The right panel shows the corresponding interactive debugging interface with labeled components: (A) a query input area, (B) retriever configuration with chunk size controls and retrieved chunk visualization, (C) LLM prompt editing area with response display, and (D) answer output with evaluation options. The interface includes interactive buttons for running the pipeline, editing retrieval parameters, testing individual steps, selecting chunks, visualizing chunk scores, editing prompts, and creating ground-truth datasets through a "save answer" feature. This interface enables developers to inspect and modify each component of their RAG pipeline while observing how changes affect intermediate outputs and final results.}
\caption{Python implementation of a RAG pipeline (left) and the corresponding debugging interface generated by \sysname (right). \sysname dynamically generates cells for each primitive component: (A) Query, (B) Retriever, (C) LLM, and (D) Answer---enabling developers to view and manipulate intermediate outputs (e.g., chunks (B2, B3), LLM responses), tweak parameters (e.g., chunk size (B1), prompts (C1)), and build a ground-truth dataset by saving answers (D1).}
\label{fig:interfacewithcode}
\vspace{-10pt}
\end{figure*}

\section{Introduction}
\label{sec:intro}

The promise of AI assistants that can engage in natural conversations with end-users, while drawing on extensive proprietary knowledge, has captured the imagination of businesses and developers alike. People, in organizations big and small, now want to build AI assistants that reliably answer questions by combining state-of-the-art general-purpose LLMs (e.g., OpenAI's gpt-4o) with proprietary organizational data---such as documents, databases, logs, and spreadsheets---containing information the models were not trained on. The de facto paradigm for building such assistants is {\em retrieval-augmented generation} (RAG)~\cite{lewis2020retrieval}. Recent surveys indicate its widespread adoption; in 2024, 86\% of enterprises deploying LLMs reportedly used RAG~\cite{k2view2024genai}.

RAG pipelines typically involve dividing documents in a corpus into ``chunks,'' or segments of text that can be individually {\em indexed} in an offline phase, and then {\em retrieved} at query time (i.e., online)~\cite{lewis2020retrieval}. Indexing creates searchable representations of these chunks---often as vector embeddings or keyword-based structures---that enable efficient retrieval. For example, a medical center might index thousands of pages of clinical guidelines. Then, in the online phase, queries can leverage the index. For example, when a patient asks {\em ``What medications should be avoided during pregnancy?,''} the pipeline retrieves some number of chunks most similar to the query, then passes these chunks to an LLM to generate a comprehensive answer. In the simplest RAG pipeline, a user query triggers one retrieval of relevant chunks via keyword or embedding-based  search, which are then incorporated into one LLM prompt to generate a response, but multiple steps of retrieval and generation are common.

Unfortunately, building effective RAG pipelines is difficult for developers. A core challenge is ensuring quality across both retrieval and LLM generation components: balancing relevance, correctness, and conciseness of answers~\cite{yang2025ragva, barnett2024seven, parnin2023building,wu2024faithful} while evaluating with only a limited ``training'' set of queries~\cite{yang2025ragva, parnin2023building, mao-etal-2024-rag}. This challenge is compounded by the fact that retrieval and LLM generation component accuracies depend on each other, making it difficult to isolate which components are causing the downstream error(s) and why. To debug these issues, developers face an extensive parameter space: in retrieval, they can tune chunk granularity, overlap, representation methods (e.g., embeddings, term frequency vectors), and the number of retrieved chunks~\cite{sivasothy2024ragprobe}; in LLM generation components, they need to optimize prompt instructions; and at the pipeline level, they must determine the composition and sequence of retrieval and generation steps. In fact, a number of sophisticated RAG pipeline architectures exist~\cite{ma2023query, lee2022generative, jin2024long, asai2023self}, with any number of retrieval and LLM generation steps, in any order. Developers' ability to explore this vast parameter space is limited by prohibitively slow iteration cycles---changing parameters often requires re-indexing documents, which can take hours~\cite{akkiraju2024facts}. Overall, developers lack integrated tools that address this twofold challenge: the deeply intertwined nature of RAG components, and the need for low-latency experimentation, across a vast parameter space to jointly optimize retrieval and generation quality, as well as that of the overall pipeline.

In this paper, we investigate how interface design can better support the unique challenges of RAG pipeline development. We propose \sysname, a RAG developer tool that combines a Python library of composable primitives for indexing, retrieval, and LLM-based generation with an interactive web-based interface that enables real-time testing and iteration. As shown in \Cref{fig:interfacewithcode}, developers write code using our primitives to wrap RAG components. When they execute their pipeline on a query, an interactive interface appears in their browser, where they can modify parameters and explore ``what-if'' scenarios in real time. Based on insights from a formative study with experts, we designed \sysname to provide three key capabilities: first, specialized debugging visualizations tailored to both retrieval and LLM components that show distributions of outputs (such as chunk similarity scores), highlighting potential failure modes. Second, a low-latency ``what-if'' analysis framework that allows developers to dynamically modify retrievers (e.g., keyword search or embedding similarity), swap or add new pipeline components, and adjust parameters (e.g., prompts, chunk sizes) while immediately observing effects on specific queries. Third, a persistent test suite that allows developers to incrementally build a collection of representative queries, enabling consistent evaluation as the pipeline evolves.

Then, to understand whether \sysname facilitates RAG pipeline development and how experienced developers approach RAG challenges, we conducted a user study with 12 engineers who have built production-grade RAG pipelines. Participants completed think-aloud sessions while developing a hospital documentation question-answering system based on real-world medical guidelines from \papertext{a real hospital system (name omitted)}\techreport{West Virginia University Health System}. Our qualitative analysis revealed that developers consistently debug retrieval components first, before investigating LLM components, even when an LLM component precedes the retrieval component or when outputs suggest LLM reasoning errors. Our findings also reveal how developers continuously alternate between understanding documents, queries, and pipeline architecture: for example, when discovering limitations in retrieved chunks, participants typically increased chunk size to capture more context, which required adjusting LLM prompts to handle expanded input (e.g., explicitly instructing the LLM to focus on relevant information). Similarly, adding query rewriting steps required developers to reconsider retrieval component parameters. {\bf \em On average, 71.3\% of parameter changes made by each participant would have required time-consuming re-indexing in traditional workflows}---underscoring why \sysname's ability to eliminate this latency was particularly valuable for maintaining development momentum.

Overall, we contribute: {\em (i)} the design and implementation of \sysname, a specialized interface for RAG development that supports rapid iteration across pipeline components; {\em (ii)} a qualitative evaluation revealing expert debugging patterns and how developers engage in simultaneous data, retrieval, and LLM sensemaking; and {\em (iii)} design implications for future RAG tools that better align with developers' natural workflows.

\section{Related Work}
\label{sec:related}

Our work builds on previous research in debugging interfaces and addresses specific challenges in RAG pipeline development. We first review debugging work, then examine the technical challenges unique to RAG pipelines.

\subsection{Interfaces for Debugging}

Effective debugging interfaces allow developers to understand system behavior, test hypotheses, and iteratively refine implementations. Below, we discuss interfaces across software, search, LLM, and ML domains that inform our work.

\topic{General Software Debugging Interfaces} Software debugging tools traditionally support runtime inspection through features like breakpoints, variable inspection, and call stack views~\cite{ko2008debugging}.  When developers are uncertain about program behavior---a common situation with complex systems like RAG pipelines---debugging strategies often focus on answering ``why'' questions. Interfaces like WhyLine~\cite{ko2004designing} allow developers to ask ``why'' and ``why not'' questions. Other strategies---visualizing hidden program states~\cite{oney2014interstate}, enabling rapid iteration~\cite{bragdon2010code}, and linking code to effects~\cite{lieber2013theseus}---further aid understanding. While these principles are foundational, RAG debugging requires adaptation to address challenges in natural language reasoning and interactions between retrieval and generation, rather than traditional program logic.

\topic{Information Retrieval (IR)} IR systems operate in two phases: during indexing, documents are chunked and represented using either traditional statistical methods like TF-IDF~\cite{ramos2003using} or modern neural embeddings~\cite{karpukhin2020dense}, then stored for efficient lookup~\cite{manning2009introduction}. At query time, a scoring function ranks each chunk by relevance to the input.
While traditional IR debugging focuses on tools that visualize feature importance and score distributions across retrieved results~\cite{hearst2009search, choo2018visual, hohman2018visual}, RAG pipelines present fundamentally different challenges. 
Unlike traditional IR, where humans consume the results, RAG pipelines pass retrieved content to black-box LLMs, which use it in complex and unpredictable ways. For example, developers must choose retrieval methods, chunk sizes, and how many results to include in the prompt---decisions that often require trial-and-error. Our interface visualizes both retrieval quality and its impact on LLM outputs to support this process.

\topic{LLM Pipeline Development Interfaces} LLMs enable complex applications like code generation and storytelling~\cite{mann2020language, swanson2021story, gero2023social}, but present challenges such as hallucinations and unexpected reasoning failures~\cite{zamfirescu2023johnny}. Interfaces like PromptChainer and ChainForge~\cite{wu2022promptchainer, arawjo2023chainforge} help users build and test LLM chains---sequences of LLM calls---with support for prompt engineering. Tools such as PromptSource~\cite{bach2022promptsource} and methods like interactive visualizations~\cite{strobelt2022interactive} enable real-time comparisons of prompt variants. Some tools are explicitly designed for LLM output sensemaking, through organizing and comparing multiple outputs~\cite{gero2024supporting, suh2023sensecape, almeda2024prompting, huh2023genassist}. Our work extends these techniques to better support RAG pipelines, which require attention not just to prompts, but also to retrieval quality and its downstream effects. 

\topic{Interactive Tools for Multi-Component Systems}  Debugging systems with multiple components, like ML models, retrievers, and other data transformations, requires specialized tools~\cite{ono2020pipelineprofiler, hohman2024talaria}. Prior to LLMs, interfaces for debugging traditional ML applications supported patterns such as ``what-if'' parameter testing~\cite{wexler2019if}, component dependency visualization~\cite{kahng2017cti}, and linking UI elements to pipeline steps~\cite{bauerle2022symphony}. Human-in-the-loop tools also enabled interactive model tuning and data exploration~\cite{patel2010gestalt, amershi2014power}. LLMs introduce new challenges: outputs are non-deterministic, and debugging involves interpreting natural language outputs, unlike the numerical data outputs given by traditional ML systems. Recent tools for LLM-based multi-agent systems---such as AGDebugger~\cite{epperson2025interactive}, LangGraph~\cite{langgraph2024}, and AutoGen~\cite{dibia2024autogen}---support execution breakpoints and mid-run message edits, but lack  support for document retrieval. A gap thus remains in debugging tightly-coupled retrieval-generation workflows.

\subsection{RAG Pipeline Development}
\label{sec:related-ragchallenges}

We review core components, architectures, and challenges in building RAG pipelines.

\topic{Indexing and Retrieval} RAG pipelines begin by preprocessing documents into text ``chunks.'' Chunk size is a key design choice: small chunks risk losing context, while large ones may include irrelevant information or exceed LLM input limits. These chunks are then {\em indexed} offline using keyword-based methods like TF-IDF~\cite{ramos2003using}, which score word importance based on frequency and rarity, or neural embeddings~\cite{reimers2019sentence}, which map text into high-dimensional vectors using embedding models like OpenAI's or SBERT~\cite{reimers2019sentence}. While embeddings capture semantic similarity, they often lose domain-specific details and struggle with long texts~\cite{zhang2019biowordvec, zhou2024length}. Many pipelines adopt hybrid approaches that combine traditional and neural techniques. For example, Max Marginal Relevance~\cite{carbonell1998use} selects chunks iteratively, choosing each new chunk based on a weighted combination of its similarity to the query and its dissimilarity to already-selected chunks. There are also new LLM-specific techniques---RAPTOR~\cite{sarthi2024raptor} takes a different approach by not just directly embedding chunks, but also recursively clustering text fragments and generating summaries of clusters to build a hierarchical tree, enabling retrieval at different levels of abstraction across lengthy documents. Methods might also combine keyword search with embedding-based approaches~\cite{lewis2020retrieval}. Each retrieval method involves several configurable parameters, and there is rarely an obvious optimal choice for a given application~\cite{bornea2024telco}.

\topic{RAG Pipeline Architectures} The simplest RAG pipeline---a single retrieval step followed by generation---often struggles with complex information needs. To address these limitations, more sophisticated architectures have emerged.
First, {\em query rewriting} uses an LLM to reformulate ambiguous questions to be more amenable to retrieval \cite{ma2023query}. Another strategy is {\em multi-hop retrieval}, used when answers require connecting information across different documents or chunks~\cite{lee2022generative}. Third, {\em re-ranking techniques} attempt to improve retrieval precision by applying additional LLM steps \cite{jin2024long}, {\em before} generating the answer. Before generation, an LLM can evaluate and reorder retrieved chunks based on their actual relevance to the query, compensating for limitations in embedding-based similarity. Or, after generation, an additional {\em answer refinement} LLM step can verify initial LLM-generated responses against the retrieved information \cite{asai2023self}, identifying and correcting potential inaccuracies to reduce hallucinations. More broadly, a pipeline can include any combination and sequence of retrieval and LLM-based generation steps. The best architecture for an application must be discovered through experimentation.

\topic{RAG Development Challenges} Our understanding of RAG challenges largely comes from case studies and interviews. \citet{barnett2024seven} identify seven failure modes, including missing content, poor retrieval ranking, context truncation, and extraction errors. Domain-specific work, like \citet{bornea2024telco} in telecommunications, shows how structure in documents (e.g., hierarchies) shapes chunking and retrieval strategies. To address these challenges, researchers have proposed evaluation frameworks. RAGGED~\cite{hsia2024ragged} automates hyperparameter tuning across retrievers, embedding models, LLMs, chunk sizes, and context limits. RAGAS~\cite{es2023ragas} and RagProbe~\cite{sivasothy2024ragprobe} offer metrics for faithfulness, relevance, and context precision. However, these tools have practical limitations: optimal configurations are often unknown, labeled data is scarce, and the range of possible queries is vast and unpredictable~\cite{parnin2023building, akkiraju2024facts, barnett2024seven}. As a result, robust RAG pipelines tend to emerge through iterative, developer-driven refinement.

While prior work identifies common RAG challenges, it offers limited insight into how developers navigate them in practice. Existing tools also lack support for rapid iteration, observability, and debugging across retrieval and LLM components. We build on this work with a different methodological approach: using our interactive tool as a design probe to study real-world RAG pipeline debugging. Through this, we contribute both the tool itself and new insights into the strategies developers use and the challenges they face when building RAG pipelines.

\section{Formative Study \& Design Goals}
\label{sec:design}

Before articulating our design goals, with IRB approval, we conducted formative interviews with 6 practitioners (P1-P6) experienced with RAG in production. Participants included CTOs, founders, engineers and researchers across finance, healthcare, and enterprise software. We used purposive sampling~\cite{patton2002qualitative} to identify participants from our networks and public developer meetups. Interviews lasted 60-90 minutes, exploring experiences building and evaluating RAG pipelines. We took notes during interviews and did thematic analysis to identify common pain points. Our script is detailed in \Cref{app:formative-interview}. We subsequently describe the challenges we found.

First, we uncovered {\bf indexing and chunking difficulties}. All participants identified document processing, particularly chunking, as a significant challenge. P1 emphasized that \blockquote{there is not a one-size-fits-all chunk size} and that \blockquote{splitting by a certain chunk size leads to context being leaked.} P5 noted each document set required different approaches to create \blockquote{semantically coherent chunks,} highlighting scalability challenges. P3 considered indexing \blockquote{the most complex part of the system} and was interested in automated ways to determine chunk size. Existing tooling (e.g., LangChain, LlamaIndex) was deemed insufficient \blockquote{out-of-the-box} (P3, P5). These experiences underscore the need for more sophisticated chunking methods, while accommodating different document structures.

Second, we uncovered {\bf evaluation challenges with limited data}. P3 admitted to not having \blockquote{a method of evaluation beyond their gut feeling,} while P2 described their RAG pipeline's maturity as \blockquote{2/10} since they \blockquote{do not have many procedures in place for evaluation.} Even when attempted, evaluation was often ad-hoc and subjective, relying on metrics like \blockquote{how much do we like the answer?} P6's team addressed this need by asking stakeholders to specify expected queries, reference documents, and responses to construct tests. Despite awareness of evaluation tools, participants found implementing them to be \blockquote{a non-trivial amount of work} (P2). Many pipelines evolved primarily through direct user feedback rather than systematic testing.

Third, we uncovered {\bf delayed feedback cycles}. A recurring challenge was the inability to quickly test pipeline changes, leading to delayed feedback cycles. Vector databases, typically used for retrieval by embedding-based similarity, require users to define a chunk size and overlap before indexing documents, so none of our 6 participants had changed these parameters while developing pipelines due to latency. Even at query time, P3 reported frustration with pipeline latency---they wanted to disable LLM calls to speed up development. P1 noted how \blockquote{errors between [steps] can compound} when using multiple LLM calls, requiring significant time to locate sources of error. P4 mentioned that evaluating a RAG pipeline \blockquote{sometimes doesn't make sense when their prompt is constantly changing.} 
This inability to quickly test configurations led to reliance on intuition rather than systematic evaluation.

Overall, we identified 3 goals for a RAG pipeline development tool:

\begin{itemize}[nosep, left=0pt]
\item \textbf{D1: Enable rapid exploration of different pipeline configurations.} Practitioners struggle with slow feedback cycles that impede their ability to iterate on their RAG pipelines. Our tool needs to support on-the-fly testing of different parameters and configurations, with reasonable latencies (Challenge 1 and 3).
\item \textbf{D2: Support systematic evaluation with limited data.} Help practitioners build, maintain, and expand test query sets during both development and debugging (Challenge 2).
\item \textbf{D3: Integrate into engineers' existing workflows.} Given the extensive experimentation required for RAG pipeline development, our tool should fit within practitioners' Python-based ecosystem to minimize context switching.
\end{itemize}

\section{RAGGY}
\label{sec:system}

Here, we present \sysname, our tool for developing and debugging RAG pipelines. \sysname provides developers with a Python library for building RAG pipelines along with an interactive debugging interface. Developers write their RAG pipeline using \sysname's primitive components, and when executed, the system launches a React-based interactive debugging interface in the browser, connected to a Flask server that runs alongside their Python environment. In the following subsections, we describe \sysname's interactive debugging interface (\Cref{sec:system-interface}) and explain how \sysname's backend enables low-latency ``what-if'' analysis by pre-materializing retrieval indexes and checkpoints that users may want to restart pipelines from (\Cref{sec:system-implementation}).

\subsection{Debugging Interface}
\label{sec:system-interface}

\begin{figure}
\centering
\includegraphics[width=0.8\linewidth]{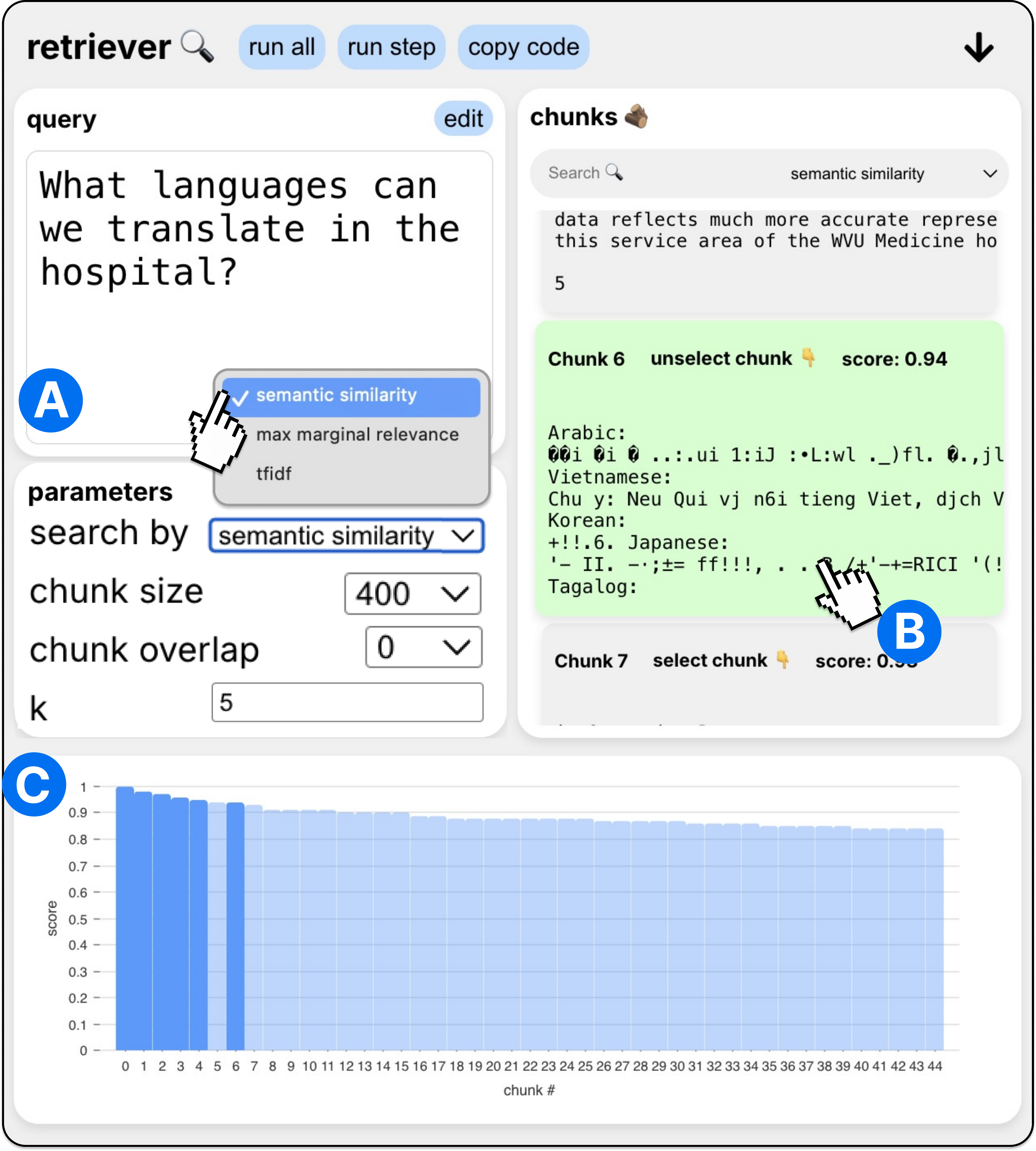}
\Description[RAGGY's interactive Retriever cell interface]{The figure shows RAGGY's Retriever cell interface. The interface is a rectangular panel with multiple interactive components:  (A) At the top left is a displayed query ``What languages can we translate in the hospital?'' is, with a parameters drop-down allowing users to configure retrieval settings including retrieval method (set to ``semantic similarity''), chunk size (set to 400), chunk overlap (set to 0), and k value (set to 5). (B) In the top-right portion, there are individual chunk displays with options to select or deselect specific chunks. Each chunk shows its content, similarity score, and has interactive controls. (C) At the bottom of the interface is a bar chart visualization showing the distribution of chunk similarity scores, with blue bars of varying heights representing different chunks. This visualization helps users understand the relative relevance of chunks and make informed decisions about which to include in the RAG pipeline.}
\caption{In \sysname's {\tt Retriever} cell, users can (A) adjust hyper-parameters (e.g., retrieval method, chunk size), (B) manually select or deselect chunks, and (C) visualize the distribution of selected vs. unselected chunks.}
\label{fig:retriever}
\vspace{-10pt}
\end{figure}

As shown in \Cref{fig:interfacewithcode}, \sysname's debugging interface provides four distinct cell types that correspond to primitive components of RAG pipelines: {\tt Query}, {\tt Retriever}, {\tt LLM}, and {\tt Answer}. We will define and describe all four types in the following paragraphs. We chose these specific primitives because, based on our formative interviews, they cover the intervention points where developers typically need to inspect and modify their RAG pipelines during debugging. 

Developers write Python scripts to implement their RAG pipelines, combining our primitives with their own logic or data processing code. They might define a list of potential queries, initialize a retriever with a collection of documents, invoke the retriever with a specific query, apply transformations to the retrieved content, pass it to an LLM, and then present the answer. While developers can include any arbitrary Python code in their scripts, only usage of a \sysname primitive automatically generates a corresponding cell in the debugging interface. When a developer runs their code, the interface dynamically populates with debugging cells in the order the steps executed (\Cref{fig:interfacewithcode}). Each cell contains specialized visualizations and controls tailored to that component type, towards our goal of enabling rapid exploration {\bf (D1)} while maintaining integration with existing workflows, described next {\bf (D2)}.

\topic{Query} In code, developers use the {\tt Query} primitive to wrap query strings in their pipeline. The {\tt Query} cell in the debugging interface (\Cref{fig:interfacewithcode}A) displays the user query and serves as the entry point to the RAG pipeline. Users can edit the query text directly in the interface and re-run the entire pipeline using the ``run pipeline'' button. While the {\tt Query} primitive essentially performs a no-op transformation in the code (i.e., it simply wraps the query string), it exists because developers frequently need to test slight variations of queries in the interface without modifying source code.

\topic{Retriever} In code, developers use the {\tt Retriever} primitive to initialize a retrieval engine with a collection of documents and invoke it with parameters such as the query, number of chunks to retrieve, chunk size, chunk overlap, and retrieval method. \sysname supports various retrieval methods including embedding-based cosine similarity, max marginal relevance, TF-IDF, and RAPTOR (as described in \Cref{sec:related-ragchallenges}). In the debugging interface, the {\tt Retriever} cell (\Cref{fig:interfacewithcode}B) provides  visualizations of similarity scores and parameters for retrieval. It displays the query text used for retrieval, a histogram visualization of similarity scores (Figure \ref{fig:retriever}C) for all chunks in the corpus (highlighting which chunks were selected), and a searchable, scrollable selector that displays all chunks sorted by relevance. This visualization helps developers quickly identify potential issues, such as when most chunks have similar scores, indicating poor differentiation. Developers can select or deselect any chunks manually, overriding the automatic retrieval process, and modify any of the retrieval parameters interactively, even switching between different retrieval methods to compare results (Figure \ref{fig:retriever}B). The cell's ``run step'' button applies these changes locally, while the ``run all'' button propagates the changes through the rest of the pipeline, with the implementation details of how \sysname achieves these interactive latencies covered in \Cref{sec:system-implementation}.

\begin{figure}
\centering
\includegraphics[width=0.85\linewidth]{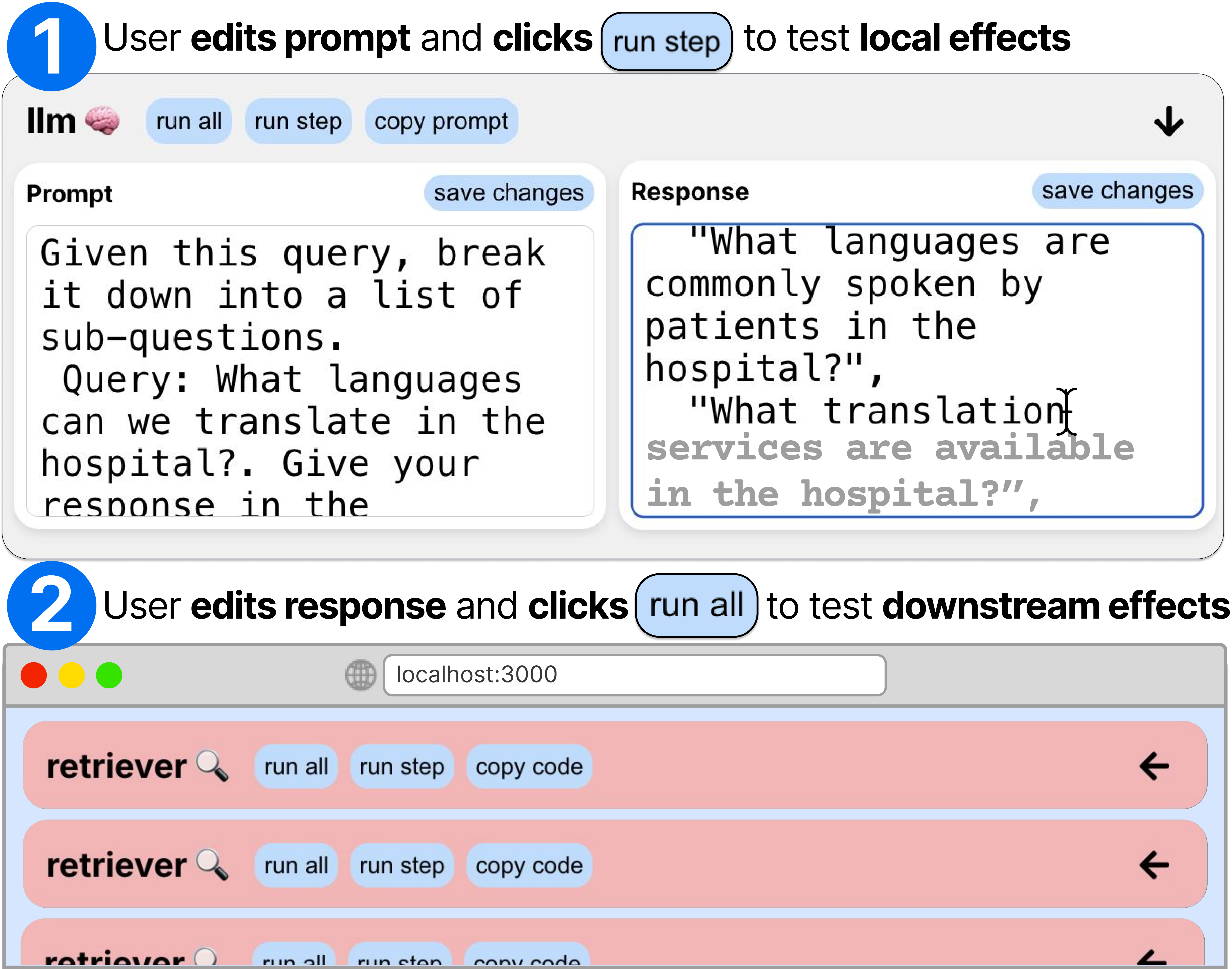}
\Description[RAGGY's interactive LLM cell interface showing query decomposition]{The figure shows RAGGY's LLM cell interface demonstrating query decomposition and response editing functionality. The interface has two numbered interaction points: (1) At the top, the user can edit prompts with controls labeled ``run step'' and ``test prompt'' alongside a prompt text area. This particular prompt details instructions for the LLM to break the query down into a list of sub-questions. The corresponding response shows the query ``What languages are commonly spoken by patients in the hospital?'' decomposed into sub-questions about translation services. (2) Below this, the user can edit the LLM's response and click ``run all'' to test how changes affect downstream components, with a series of retriever cells visible underneath that will process each sub-question. The interface uses color-coding and directional arrows to show the flow of information through the pipeline, illustrating how manipulating the LLM's query decomposition directly impacts subsequent retrieval operations.}
\caption{In \sysname's {\tt LLM} cell, users can (1) experiment with prompting strategies and (2) directly modify the LLM’s output to test downstream effects. This example shows query decomposition, where a prompt is broken into sub-questions, each triggering its own retrieval.}
\label{fig:LLM}
\vspace{-15pt}
\end{figure}

\topic{LLM Generator} In code, developers use the {\tt LLM} primitive to generate an answer a prompt, which typically incorporates retrieved context, query, and outputs from other previous components in the pipeline. The {\tt LLM} cell in the interface (\Cref{fig:interfacewithcode}C) displays the prompt sent to the language model and the generated output. The visualization shows the prompt structure, variables, and the resulting generation. Developers can edit the prompt text directly, and test different modifications to their prompt, and even edit the output text directly (Figure \ref{fig:LLM}.2). Direct manipulation of the output is particularly valuable when developers want to test how downstream components would behave with a ``perfect'' LLM response, or if they want to explore the effects of alternative phrasings of the response without waiting for additional model calls. Similar to the {\tt Retriever} cell, the ``run step'' button applies changes locally, while ``run all'' propagates changes through downstream components.

\begin{figure}
\centering
\includegraphics[width=0.8\linewidth]{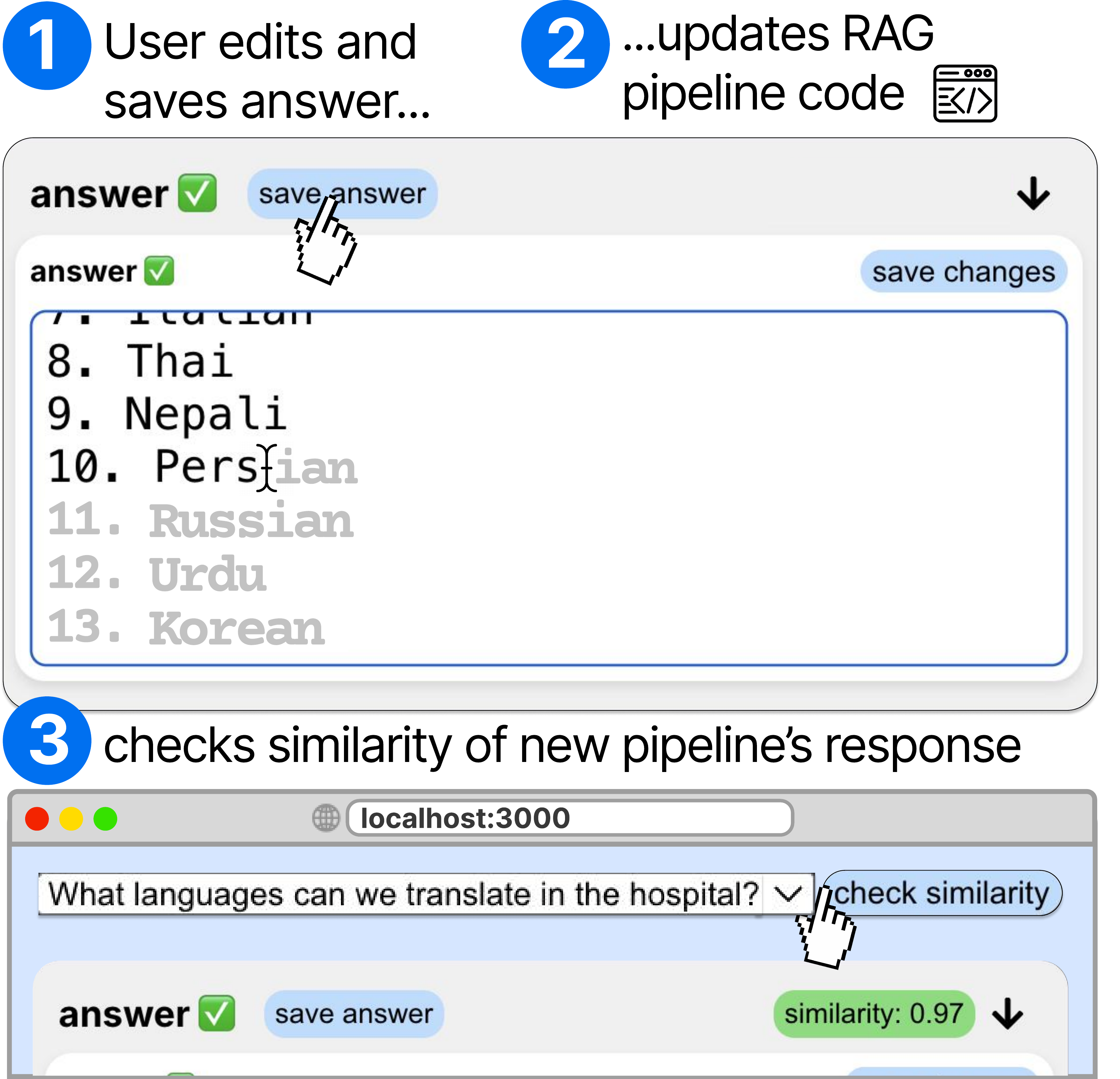}
\Description[RAGGY's Answer interface showing ground-truth comparison functionality]{The figure illustrates RAGGY's Answer interface with three numbered components demonstrating the ground-truth comparison workflow: (1) In the top section, a user is shown editing and saving an answer to a question of ``What languages can we translate in the hospital?,'' with a checkmark indicating a verified response and a cursor hovering over a ``save answer'' button. The answer content shows a numbered list of languages (including "Thai", "Nepali", "Farsi", with additional languages like "Russian", "Urdu", and "Korean" appearing grayed out). (2) An arrow indicates that saving this answer updates the RAG pipeline code, with a code icon shown. (3) In the bottom section, the interface shows how users can check similarity between new pipeline results and the saved ground-truth. This includes the original query "What languages can we translate in the hospital?" alongside a "check similarity" button being clicked, and a similarity score of 0.97 displayed. The overall flow demonstrates how users can establish reference answers and evaluate subsequent pipeline modifications against this ground-truth.}
\caption{Answer Interface. The {\tt Answer} cell comparing a current pipeline result against a previously saved ``golden'' reference. With \sysname, users can save and edit answers (1) and later check the similarity of new results from the pipeline against their saved ground-truth (3).}
\label{fig:answer}
\vspace{-10pt}
\end{figure}

\topic{Answer} In code, developers use the {\tt Answer} primitive to wrap the final response. The {\tt Answer} cell  (\Cref{fig:interfacewithcode}D) displays the final output of the RAG pipeline and provides functionality evaluation. Developers can view and edit the generated answer text directly, and save the current answer as a ``golden'' reference using the ``save answer'' button (\Cref{fig:answer}.3). When saved answers exist for a query, the {\tt Answer} cell also displays a similarity score (\Cref{fig:answer}.3) between the current answer and the saved golden answer (e.g., the cosine similarity between the embeddings of the answer), helping developers track how pipeline changes affect output quality {\bf (D3)}. \Cref{fig:answer} shows an example of an answer's similarity with a golden reference.

\begin{figure*}
\centering
\includegraphics[width=0.8\linewidth]{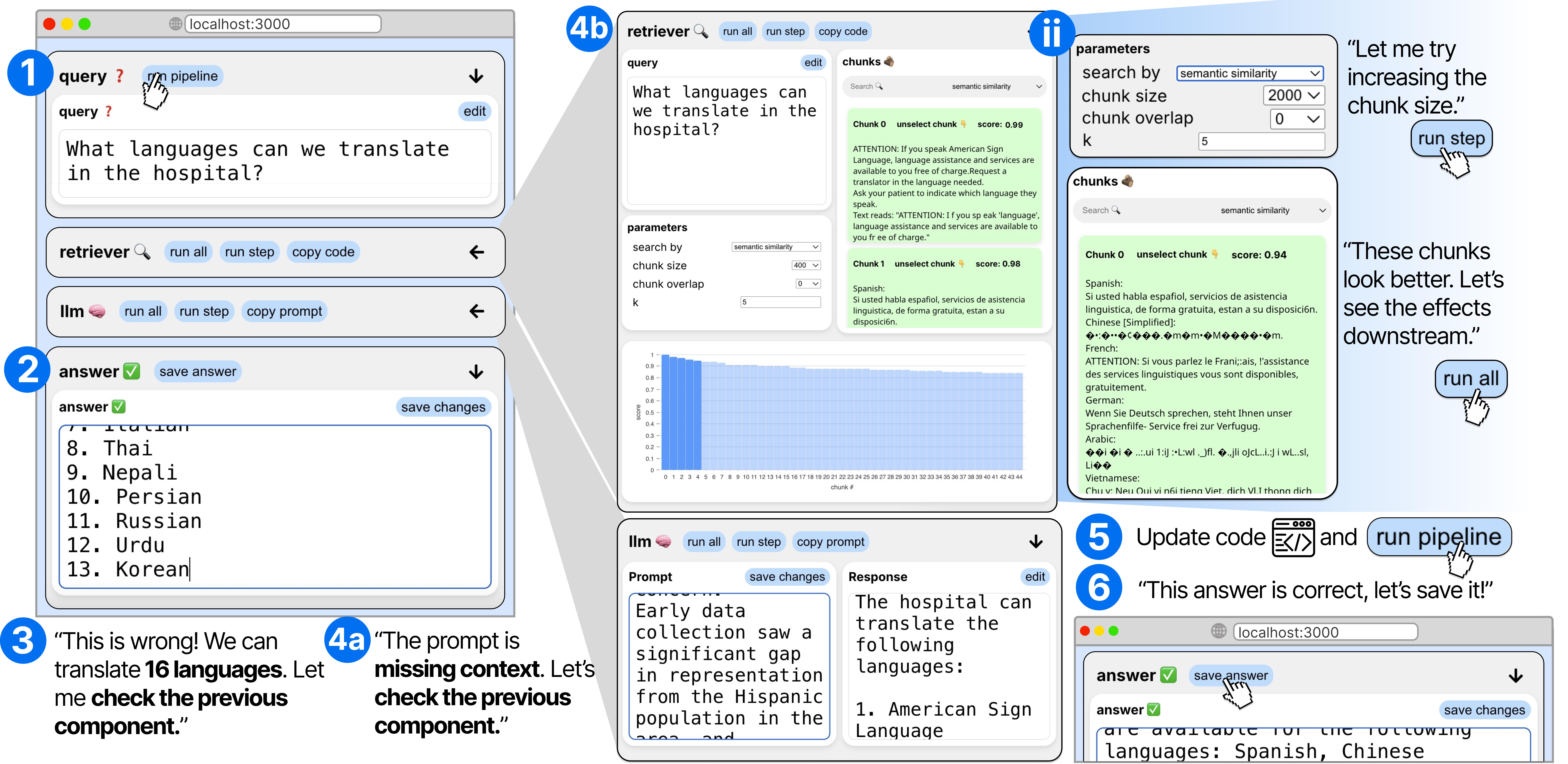}
\Description[RAGGY's iterative debugging workflow across multiple interface components]{The figure illustrates RAGGY's complete iterative debugging workflow through multiple interconnected interface panels. The workflow proceeds in numbered steps: (1) Upper left shows a user running the initial pipeline with the query ``What languages can we translate in the hospital?''. (2) Below this, the answer panel displays a numbered list of languages including Thai, Nepali, Persian, Russian, Urdu, and Korean. (3) Speech bubbles indicate user evaluation, stating ``This is wrong! We can translate 16 languages. Let me check the previous component.'' (4) The debugging branches into two paths: (4a) LLM debugging where the user edits prompts that are "missing context" and (4b) retriever debugging showing the detailed retriever interface with parameter controls and chunk visualization. A speech bubble states ``Let me try increasing the chunk size.'' The right side shows: (i) Interactive testing with a ``run step'' button, (ii) A downstream test with ``These chunks look better. Let's see the effects downstream'' and a ``run all'' button, (5) Code update functionality with an update icon, and (6) Evaluation building with the text ``This answer is correct, let's save it!'' and a final answer panel showing translation capabilities for American Sign Language, Spanish, and Chinese. The entire layout demonstrates how users can identify issues, test solutions at different pipeline stages, and progressively improve their RAG system while building an evaluation dataset.}
\caption{The iterative debugging workflow supported by \sysname. Users can (1) run the pipeline, (2) inspect the answer, and (3) backtrack if it’s incorrect. They can then (4a) debug the LLM by editing prompts or parameters, or (4b) debug the retriever by examining retrieved chunks and tuning retrieval settings (e.g., chunk size, method). After testing changes locally and downstream (ii), users can (5) update code, (6) and save correct answers to iteratively build an evaluation set. Finally users can repeat steps 1-6 as needed to refine the pipeline.}
  \label{fig:debugging-workflow}
\vspace{-10pt}
\end{figure*}

By dynamically generating the interface at pipeline runtime, \sysname supports debugging complex pipelines, allowing developers to use loops, conditionals, and any sequence of primitive components in their code. The interface displays the expanded execution plan for a given query. Additionally, to minimize friction when switching between code and the debugging interface~\cite{ko2004designing}, \sysname includes a ``copy code'' button that generates code for a given cell based on its parameters.
All together, \Cref{fig:debugging-workflow} details the iterative debugging workflow \sysname supports.

\subsection{Backend Implementation}
\label{sec:system-implementation}

\sysname enables low-latency ``what-if'' queries, or changes developers might make in the debugging interface, through {\em pre-computed vector indexes} and {\em pipeline-specific program checkpoints}. 

\subsubsection{Pre-computed Vector Indexes for Retrieval} When a developer first runs a \sysname pipeline, the system executes a one-time preprocessing operation to create several indexes for their document corpus. We use a Chroma database to store our vector indexes~\cite{chroma2025}. \sysname automatically generates different chunking configurations, dividing documents into chunks with various combinations of chunk sizes (100-2000 characters) and chunk overlaps (0-400 characters). 
Chunk sizes were selected at increasing intervals, with finer granularity at lower values to capture variations in small-document segmentation and larger gaps at higher values to balance efficiency~\cite{kaggle_chunk_guide}. 
Chunk overlaps were chosen to provide a mix of different redundancy configurations~\cite{kaggle_chunk_guide}.
For each chunking configuration, \sysname creates four different indexes: {\em (i)} cosine similarity-based retrieval, {\em (ii)} TF-IDF frequency-based retrieval, {\em (iii)} max marginal relevance-based retrieval, and {\em (iv)} RAPTOR (i.e., embeddings of summaries of chunks). Indexing options {\em (i)}, {\em (iii)}, and {\em (iv)} leverage OpenAI's \ttt{text-embedding-3-large} model. Overall, this results in hundreds of pre-computed indexes that cover a large portion of the parameter space developers typically explore during RAG optimization, and can take over an hour (though this is a one-time cost).

With these indexes pre-materialized, subsequent pipeline runs or interactive modifications can immediately leverage the appropriate index without waiting for document reprocessing. For example, when a developer adjusts the chunk size from 200 to 500 characters in the \texttt{Retriever} cell, \sysname instantly queries the corresponding pre-computed index rather than re-processing the entire document collection---reducing what would typically be minutes (or hours) of processing to under a second.

\subsubsection{Program State Preservation}

To enable users to modify components mid-execution in the debugging interface, \sysname needs to ``go back'' to the program state at the time when specific primitive components were executed. \sysname creates checkpoints at each primitive component invocation; when a primitive component (e.g., {\tt Retriever} or {\tt LLM}) executes, \sysname forks the Python process before the component returns its value, creating a copy of the program's state at that exact moment. The parent process continues normal execution, while the child process is paused and registered with the debugging server. 

When a user modifies parameters in the debugging interface and clicks ``run,'' the server signals the appropriate child process to resume. At this point, the child process takes over as the new ``main'' process for the pipeline execution, and the original parent process terminates. As this new main process continues execution, it creates its own checkpoints at subsequent primitive components, forking new child processes that register with the server. To prevent resource exhaustion from accumulating sleeping processes, \sysname cleans them up as follows: when a parent process terminates after its child takes over execution, the debugging server identifies and terminates all sleeping child processes that were registered by the now-terminated parent. Only the currently active execution path and its immediate checkpoints remain in memory, effectively allowing for unlimited interaction cycles without memory leaks.

\section{User Study Design}
\label{sec:userstudy}

To understand how practitioners use \sysname and learn more about expert RAG pipeline development strategies, we conducted a user study with 12 participants experienced in RAG pipeline development. Participants completed programming tasks using \sysname while thinking aloud.

\topic{Recruitment, Study Participants, and Procedure} We recruited 12 participants through ML/LLM communities and the authors’ networks. All had Python proficiency and prior experience deploying RAG pipelines. Participants ranged from students with internship experience to professionals who built large-scale ``AI copilots'' for millions of users (\Cref{tab:participants}). Each one-hour interview included onboarding (5 min), coding tasks (45 min), and a semi-structured interview (10 min). After a demo of \sysname, participants reviewed the task description and starter code. The task involved improving a naive RAG pipeline for question-answering over a corpus of 220 PDFs from \papertext{the website of a real hospital system (name omitted)}\techreport{the West Virginia University Health System's website}, with three stages:

\begin{enumerate}[nosep, left=0pt] 
\item {\bf Stage 1:} Handle well-formed queries (e.g., ``How many beds are in [hospital name] Medical Center?''). These varied in difficulty,  including single-hop (answerable from one chunk), multi-hop (requiring information from multiple chunks), and reasoning-intensive questions. Ground-truth answers and source document links were provided. 
\item {\bf Stage 2:} Handle noisy, keyword-style queries reflecting real user behavior (e.g., ``num beds [hospital]'', ``ovr awrd polcy''). 
\item {\bf Stage 3:} Detect and reject irrelevant or off-topic queries rather than attempting to answer them. 
\end{enumerate}

The baseline pipeline used embedding-based retrieval (chunk size 200, no overlap, $k=5$) and an LLM for answer generation. This baseline configuration was informed by input from a collaborator working at the hospital system. Participants were encouraged to think aloud, explore freely, and focus on process over completion. We provided support for using the interface and a cheat sheet of common RAG design patterns (e.g., re-ranking, query decomposition; see \Cref{app:task-cheat-sheet}).

The semi-structured interview in the last 10 minutes focused on participants’ workflows with \sysname and how they compared it to their typical RAG development process (see~\Cref{app:raggy-user-study}). Sessions were conducted via Zoom, using remote control to let participants interact with \sysname and edit code on the first author’s machine. The study was IRB-approved, and all participants generously volunteered their time with no monetary compensation.

\topic{Analysis} Participants were asked to think aloud while using \sysname, as we took notes on their thoughts and visible reactions (e.g., excitement or frustration). Zoom auto-generated transcripts were corrected and annotated with interface references. We used open coding and affinity diagramming~\cite{taylor2015introduction} to identify themes across transcripts and notes. Codes were iteratively grouped into high-level themes and refined through team discussions.

We focused on qualitative insights rather than quantitative measures in our study because RAG development is a new, complex practice involving multiple interacting components (retrievers, LLMs, chunking), and standard survey instruments might be inadequate for capturing developers' detailed experiences. Our qualitative approach follows HCI best practices for studying early-stage developer workflows~\cite{ko2015practical, murphy2002evaluating, lazar2017research} and allows us to surface patterns that quantitative methods might overlook.

\section{User Study Findings}
\label{sec:findings}

Here, we describe the findings from our user study. Overall, we found the following:
\begin{enumerate}
[nosep, left=0pt]
    \item \textbf{Holistic Evaluation:} Participants valued \sysname's fast iteration but wanted more systematic evaluation capabilities.
    
    \item \textbf{High-Level Exploration Patterns:} Developers prioritized validating retrieval first, then followed diverse paths.
    
    \item \textbf{Component Investigation:} Developers engaged in iterative information foraging and sensemaking~\cite{pirolli1999information} to understand individual components.
    
    \item \textbf{Component Interactions:} Changes to one component often required adjustments to others, highlighting the need to manage interdependencies.
\end{enumerate}
We unpack these findings below.

\subsection{Meeting Design Goals and Identifying Unmet Needs}

 In this section, we detail how \sysname met our design objectives, and where opportunities for improvement emerged.

\subsubsection{``Half a Day's Work'' Becomes ``Just Switch It Out in the Dropdown''} Participants expressed enthusiasm about the system's ability to facilitate rapid pipeline iteration {\bf (D1)}, saying things like, \blockquote{I love this} (P2), \blockquote{I would love to have this running in a little docker container} (P10), \blockquote{This is awesome. This is really cool} (P6), and \blockquote{This is actually very helpful} (P3).  The enthusiasm stemmed primarily from the system's ability to significantly reduce the time needed to implement and test changes. P5 appreciated the ability to test changes incrementally: \blockquote{I think being able to run this in steps is effective for testing any changes I want to implement.} P1 echoed this sentiment: \blockquote{Being able to [change] each [retrieval hyperparameter], even parameter by parameter, is super beneficial.} When testing changes, developers found it very useful to locally test a change to a specific component (e.g., LLM or Retriever) using ``run step,'' and then ``run all'' to see how the changed parameters impacted the rest of the pipeline. P5 reflected that testing pipeline changes was previously \blockquote{like half a day's work,} whereas with \sysname, they could \blockquote{just go like, you know, switch it out in the dropdown.} 
P10 similarly appreciated the short wait times between iteration cycles: \blockquote{I really liked how fast it was... It makes it [easier] to say, okay, this part in the pipeline is where things are going maybe a little astray.} 
This feeling was well-founded: \textbf{on average, 71.3\% of parameter changes made by each participants would have required time-consuming re-indexing in traditional workflows} (see Figure \ref{fig:bar-hyperparams}).
Of all retrieval parameters (e.g., chunk size, chunk overlap, retrieval method, $k$), the most frequently adjusted was chunk size. 

\subsubsection{\sysname Doesn't Add More Friction to Workflows} P6, P8, P10, and P12 specifically praised the UI for viewing chunks, appreciating how they could easily see both retrieved and non-retrieved chunks within the interface, maintaining their flow ({\bf D3}). P10 highlighted how this visibility helped validate component interactions: \blockquote{Probably my favorite part was, I can see hey, this is what's getting pulled in.} Interestingly, P6 envisioned \sysname enabling non-technical stakeholders to participate in the iteration process through: \blockquote{non-technical people could, together with the data scientist probably, look at this and review it and say like, okay, no, those chunks suck, but like, scroll down, here's the right ones we want.}

Moreover, participants (P1, P2, P4, P6) liked that the UI was auto-generated from their code, maintaining developer flow {\bf (D3)}.  P6 explicitly praised this design decision: \blockquote{I don't have to go into the interface and create this pipeline. I create the pipeline through my pipeline, right through my code, right? And the interface creates itself based upon that? Awesome idea.} P6 elaborated that in data science work, they spend most of their time iterating in code, so the reduced overhead allowed them to maintain their workflow while gaining the benefits of interactive debugging. 

\subsubsection{Users Want More Systematic Evaluation}
\label{see-all-experiments}
While participants valued the ability to save and rerun traces, several (P3-P6, P8) expressed a clear desire for \sysname to provide more systematic support for evaluating multiple traces simultaneously {\bf (D2)}. This emerged as the most significant gap in \sysname's current capabilities. P4 said: \blockquote{I think the piece I'd be really missing here. That's necessary for workflow is an accounting mechanism where I can see all my experiments and explore the results.} P6 specifically wanted to view prompt and chunk diffs over different iterations side-by-side. This need became particularly acute when dealing with multiple experimental parameters, as P3 expressed they would have to try the cross product of all parameters.
This highlights the need for tools that also systematic evaluation across multiple configurations~\cite{wang2019human, xin2021whither}.

Moreover, despite the positive reception of \sysname's retrieval inspection capabilities, participants consistently expressed a desire for better provenance when inspecting chunks. They wanted to view which specific documents chunks came from, in addition to the chunk text itself.
P10 articulated how provenance builds trust: \blockquote{I just find it a lot more trustworthy when you say, `Hey, this is from here,' and they can actually click and verify.} Similarly, P12 stated plainly: \blockquote{I just want to see the document in full.}

Finally, participants also highlighted the importance of tracking performance changes over time, particularly for deployed systems. P5 described wanting visibility into temporal changes: \blockquote{We would also want to verify that when we change the chunk size or whatever, you can see, like the difference, the delta between what the user saw like 3 months ago, in an actual conversation with our new strategy.}  Overall, while \sysname well-supports {\em single-session} iteration, developers also need tools for longitudinal evaluation across multiple sessions and versions. This reflects the reality that RAG pipeline development is not a one-time task but an ongoing process that continues throughout the entire software application lifecycle~\cite{amershi2019software, shankar2024we}.

\subsection{Developers Focus on Retrieval First, Then Follow Diverse Paths}
\label{sec:retrieval-first}

Developers follow a consistent pattern when debugging RAG pipelines: they start by verifying retrieval quality, then follow non-linear paths afterward based on their findings.

\subsubsection{``I Don't Want to Work on Other Sh*t Until I Know I Can Retrieve the Right Documents.''} We observed that participants overwhelmingly took a retriever-first approach to pipeline validation (P1-P6, P8-P12). Rather than examining and validating the final output of the pipeline, developers first reviewed the retrieved chunks before assessing any upstream or downstream steps. P11 explicitly articulated this widespread sentiment: \blockquote{I don't wanna work on other sh*t until I know I can retrieve the right documents.} Similarly, P10 explained, \blockquote{You can iterate [on] a prompt all you want. If you're giving it bad context as data, then that's a big issue. Right?} While P10 had not previously used tools that isolated retrieval, they noted that iterating on retrieval first was most intuitive to them and expressed a desire to iterate on retrieval more in their own work after the study. Some participants went as far as copy and pasting queries into only the retriever component, instead of re-running the pipeline, as the interface design suggested (P3-P5, P8). This finding echoes the sentiment observed by \citet{ko2006exploratory}; when debugging, developers first focus their attention on the most critical, uncertain components of broader software systems.\footnote{Perhaps retrieval feels more uncertain to developers than LLMs because it depends directly on one's familiarity with the document corpus. Unlike LLMs, being able to assess retrieval quality hinges on knowing what information actually exists in the corpus---knowledge developers often lack.}

\newcolumntype{L}[1]{>{\raggedright\arraybackslash}p{#1}}

\begin{table}[t!]
\centering
\footnotesize
\begin{tabular}{@{}L{1.3cm}L{2cm}L{4cm}@{}}  %

\toprule
\textbf{PID} & \textbf{Age} & \textbf{Occupation} \\ 
\midrule
P1 & 20-30 & Software Engineer \\
P2 & 50-60 & VP \& CIO \\
P3 & 30-40 & Software Engineer \\
P4 & 30-40 & ML Engineer \\
P5 & 20-30 & Software Engineer \\
P6 & 30-40 & Data Scientist \\
P7 & 40-50 & Principal Engineer \\
P8 & 40-50 & Director - Machine Learning \\
P9 & 30-40 & Research Scientist \\
P10 & 30-40 & ML Engineer \\
P11 & 30-40 & Technical Founder \\
P12 & 30-40 & Research Scientist \\
\bottomrule
\end{tabular}
\Description[Table showing demographics of study participants]{This table presents demographic information about the 12 study participants (P1-P12). The table has three columns: Participant ID (PID), Age range, and Occupation. Participants span age ranges from 20-30 to 50-60, with most participants (7) in the 30-40 age range. Two participants are in the 20-30 range, two in the 40-50 range, and one in the 50-60 range. Occupations represent various AI and technology roles: three Software Engineers, two ML Engineers, two Research Scientists, one Data Scientist, one Principal Engineer, one Director of Machine Learning, one Technical Founder, and one VP \& CIO. This participant pool represents professionals who build and deploy production RAG applications across diverse AI-related positions.}
\caption{Study participants included professionals building and deploying production RAG applications across various AI-related roles.}
\label{tab:participants}
\vspace{-20pt}
\end{table}

Several developers (P1, P3, P6, P9, P12) performed their own systematic sanity checks on retrieved chunks---mentally evaluating retrieval precision, even without automated evaluation metrics. For instance, after inspecting a set of retrieved chunks, some participants would mention how the retrieval was \blockquote{not precise enough} (P12) or note that a target chunk \blockquote{doesn't rank highly enough} (P12). P12 described their process as \blockquote{essentially doing human keyword search} over the set of retrieved chunks. 

\subsubsection{Developers Follow a Non-Linear Paths}  Based on our observations, the typical workflow developers followed with \sysname was:

\begin{enumerate}[nosep, left=0pt]
    \item \textbf{Initial Foraging and Sensemaking:} All developers sought to form a basic understanding of the document corpus, query types, and pipeline structure, with some participants opting to read the documents to form a better understanding of the task.
    \item \textbf{Initial Pipeline Run:} Developers ran sample queries through the pipeline to identify early failure points.
    \item \textbf{Retriever Inspection:} All participants (P1–P12) examined retrieved chunks. {\bf \em Here is where participant workflows diverged.} If the chunks met expectations (P1, P2, P6, P9), they moved on. Others returned to sensemaking or changed retrieval parameters like \ttt{chunk size}. Some (P3-P5, P8) tested all queries manually at this step.

    \item \textbf{LLM Inspection:} Developers (P1–P12) reviewed the prompt and output. Many modified the prompt (P1, P3–P8, P10) or tuned LLM parameters such as \ttt{max tokens} and \ttt{temperature} (P5, P7, P11, P12). Some (P1, P5, P6) also validated how chunks were integrated into prompts.

    \item \textbf{Answer Validation:} All compared outputs to the ground truth. If correct (P1, P2, P6, P7, P12), they advanced to new queries. Some (P3-P5, P8) emphasized chunk quality over final answers, while others (P10, P11) imposed additional constraints beyond the provided ground truth.

\end{enumerate}

Essentially, after validating retrieval, developers would move between components based on their sensemaking needs---whether they needed to better understand the document set, the range of possible queries, or the interactions between pipeline components. This non-linear debugging approach mirrors findings by \citet{guo2011proactive}, who observed similar non-linear information seeking patterns in programmers' workflow during debugging tasks.\footnote{At a meta level, RAG pipeline debugging can be viewed as a form of data wrangling with an LLM in the loop---developers iteratively transform, evaluate, and refine data flows until they produce the desired outputs.} 

While participants followed similar workflows, {\em how} they interpreted the same results different, and their subsequent iterations varied. For example, when encountering incorrect chunk retrieval, P7 attributed it to poor query specificity and implemented a query re-writer, while P8 simply increased chunk size. Though P7's initial solution failed to resolve the issue, they eventually also increased chunk size after better understanding retrieval parameters. Despite taking different reasoning paths, both participants ultimately converged on similar pipeline architectures.

\subsection{Developers Engage in Iterative Foraging and Sensemaking Loops}
\label{sec:sensemaking}

\definecolor{lightblue}{RGB}{69,155,255}
\definecolor{lightgreen}{RGB}{204,255,204}
\definecolor{lightgray}{RGB}{240,240,240}
\definecolor{white}{RGB}{255,255,255}

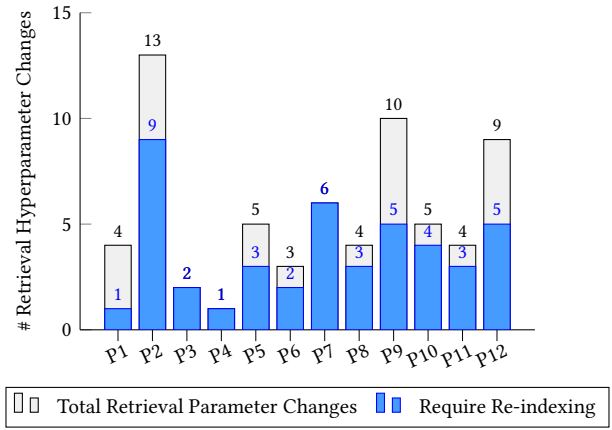
\begin{figure}[t!]
\label{parameter-changes}
\centering
\begin{tikzpicture}
\begin{axis}[
    title={\textbf{Retrieval Hyperparameter Changes Per Participant}},
    ybar,
    xticklabel style={rotate=25, anchor=east, font=\small},
    bar width=10pt,
    width=0.90\linewidth,
    x tick label style={/pgf/number format/fixed, xshift=5pt, yshift=-4pt},
    height=5.8cm,
    enlargelimits=auto,
    ymin=0,
    ymax=15,
    ylabel={\# Retrieval Hyperparameter Changes},
    ylabel style={font=\small},
    xtick=data,
    symbolic x coords={P1,P2,P3,P4,P5,P6,P7,P8,P9,P10,P11,P12},
    xtick style={draw=black},
    tick label style={font=\small},
    nodes near coords,
    nodes near coords align={vertical},
    every node near coord/.append style={font=\footnotesize},
    axis x line*=bottom,
    axis y line*=left,
    legend style={
        font=\small,
        at={(0.5,-0.18)},
        anchor=north,
        legend columns=2,
        column sep=5pt
    },
    legend cell align={left},
    bar shift=0pt,
    cycle list name=color list
]
\addplot+[fill=lightgray, draw=black, nodes near coords, every node near coord/.append style={font=\footnotesize, color=black}] 
    coordinates {(P1,4) (P2,13) (P3,2) (P4,1) (P5,5) (P6,3) (P7,6) (P8,4) (P9,10) (P10,5) (P11,4) (P12,9)};
\addplot+[fill=lightblue] 
    coordinates {(P1,1) (P2,9) (P3,2) (P4,1) (P5,3) (P6,2) (P7,6) (P8,3) (P9,5) (P10,4) (P11,3) (P12,5)};
\legend{Total Retrieval Parameter Changes, Require Re-indexing}
\end{axis}
\end{tikzpicture}
\Description[Bar chart showing retrieval hyperparameter changes per participant]{The figure shows a bar chart titled "Retrieval Parameter Changes Per Participant" with participants P1 through P12 on the x-axis and number of hyperparameter changes on the y-axis (ranging from 0 to 15). Each participant has a two-tone stacked bar: the total bar height (light gray) represents the total number of retrieval hyperparameter changes made, while the darker portion (blue) indicates changes that would have required re-indexing in traditional workflows. Participant P2 made the most changes (13 total, with 9 requiring re-indexing), followed by P9 (10 total, 5 requiring re-indexing), and P12 (9 total, 5 requiring re-indexing). Other participants made between 1-6 changes. The visualization illustrates that across all participants, a significant proportion (71.3\%) of parameter changes would have required time-consuming re-indexing without RAGGY's interactive approach.}
\caption{Darker bars indicate changes that would have required re-indexing without \sysname. On average, 71.3\% of
parameter changes made per participant would have required time-consuming re-indexing in traditional workflows.}
\label{fig:bar-hyperparams}
\vspace{-10pt}
\end{figure}

We frame users' debugging strategies through an information seeking lens, through Pirolli and Card's model of {\em foraging} and {\em sensemaking} loops \cite{pirolli1999information}. All participants started out with a ``foraging'' pass, identifying simple queries and the documents that contained the answers to these queries. Then, participants iterated between making sense of documents and LLM behavior.

\subsubsection{Developers Initially Make Sense of Documents} Most developers (P1-P3, P5-P12) initiated their workflow by systematically exploring the document corpus to collect and filter relevant information. During this sensemaking phase, developers mostly tried to understand: the distribution of document lengths (P3, P5), common structures in the documents, document attributes and metadata (P2, P3, P5, P9-P11), any inconsistencies between documents  (P2, P3, P11), where the ground-truth answers were situated in the documents (P1-P3, P5-P7, P10-P12), and any presence of duplicate information (P1, P7). P11 expressed a preference for thorough document exploration before any pipeline iteration: \blockquote{When I first [start] a project I usually spend like a week like looking at documents.} P5 and P8 noted that comprehensive data exploration is not always possible in production environments, where there might be data drift~\cite{shankar2024we}. Importantly, we observed developers frequently transitioning back to information foraging after initial sensemaking attempts. The most common trigger for shifting from sensemaking to foraging occurred after debugging a retriever, when participants were unable to quickly ascertain how ``good'' a chunk was (P1-P3, P5-P10, P12). They needed to look at that chunk in conjunction with other parts of the document, or other similar documents.

\subsubsection{Developers Build Mental Models LLM Behavior through Queries}  Developers employed several creative strategies to make sense of how LLMs responded to the domain of queries that the system should and should not support:

\begin{enumerate}[nosep, left=0pt]
    \item {\bf Creating Personas:} P3 created personas in order to generate possibly problematic queries. Beyond simulating a ``trashy user,'' P3 also informed personas based on constraints they discovered while sensemaking over the document corpus. In one query, P3 wrote \blockquote{I am 16 years old. Can I participate in the job shadowing program?} to test the robustness of LLMs that needed to ``reason,'' in addition to simply providing context to answer. 
    \item {\bf Generating edge cases:}  P2 created intentionally ambiguous queries like \blockquote{What is the dress code policy?}---knowing this would vary by employee role and hospital location.
    \item {\bf Adding Constraints:} When iterating on one of our task questions, \blockquote{Who is on the Investigational Drug Services (IDS) team?,} P11 refused to run a pipeline without gathering additional constraints for the question. P11 reflected upon how ``who'' is ambiguous---whether the question aimed to supply direct names of leadership staff or roles. While P11's pipeline output correctly aligned with the provided ground truth, to them it was still not an acceptable output. They went further and disagreed with our provided ground-truth answer, insisting that the question needed to be better specified if the pipeline were to be productionized.
\end{enumerate}

Participants emphasized ongoing sensemaking even after deployment. P5 described how their team \blockquote{goes through a random sample of user conversations every few weeks} to track query evolution, a common ML engineering practice~\cite{amershi2019software, shankar2024we}. This continuous analysis revealed useful insights about user behavior, with P5 noting \blockquote{Users see [the chatbot] as a Google Search...}, which directly influenced their implementation of hybrid keyword and semantic search.

\subsection{Changes to Retrieval Necessitate Adjustments to Generation and Vice Versa}

\begin{table}
\label{tab:debug-tactics}
\footnotesize
\begin{tabular}{@{}p{2.1cm}@{\hskip 4pt}p{4.6cm}@{\hskip 2pt}p{1.5cm}@{}}
\toprule
\textbf{Failure Mode} & \textbf{Debugging Tactics} & \textbf{Participants} \\
\midrule

\multirow{5}{=}{Incorrect chunks retrieved} & Increase chunk size & P1–P6, P8–P12 \\
& Decrease number of retrieved chunks & P1, P2, P4 \\
& Increase number of retrieved chunks & P5 \\
& Add query rewriting step \& increase chunk size & P7 \\
& Add a query de-composer & P6 \\

\midrule
Target chunks ranked too low & Increase chunk size \& implement re-ranker & P9 \\

\midrule
\multirow{2}{=}{Answer lacks specificity or is incomplete} & Increase chunk size to include more of the source document & P4, P5, P8, P12 \\
& Adjust prompt to request more comprehensive answers & P2 \\

\midrule
\multirow{2}{=}{Ambiguous queries} & Add query re-writer & P3, P5, P8, P12 \\
& Hybrid retrieval & P5 \\

\midrule
\multirow{3}{=}{Making pipeline conversationally ready} & Modify final answer prompt & P3, P5, P7, P10 \\
& Add filtering LLM call for irrelevant questions & P8 \\
& Add domain-specific grounding to prompts & P3, P5 \\

\bottomrule
\end{tabular}
\Description{This table summarizes debugging tactics used by participants for common RAG pipeline failure modes. It includes three columns: ``Failure Mode'', ``Debugging Tactics'', and ``Participants''. For ``Incorrect chunks retrieved'', tactics included increasing chunk size (P1--P6, P8--P12), decreasing or increasing the number of retrieved chunks (P1, P2, P4, P5), adding query rewriting with larger chunk size (P7), and adding a query de-composer (P6). For ``Target chunks ranked too low'', P9 increased chunk size and added a re-ranker. For ``Answer lacks specificity or is incomplete'', participants either increased chunk size to include more of the source document (P4, P5, P8, P12) or adjusted the prompt to request more complete answers (P2). For ``Ambiguous queries'', tactics included adding a query re-writer (P3, P5, P8, P12) and using hybrid retrieval (P5). For ``Making pipeline conversationally ready'', participants modified the final prompt (P3, P5, P7, P10), added a filtering LLM for irrelevant questions (P8), or grounded prompts in domain-specific language (P3, P5).}
\caption{Debugging tactics used by participants for common RAG pipeline failures. Similar bugs led to diverse strategies for fixes, revealing a broad pipeline design space and a lack of established ``best practices'' for RAG debugging.}
\vspace{-15pt}
\end{table}

Modifying one component of a RAG pipeline frequently required coordinated changes to other components. \Cref{tab:debug-tactics} illustrates common debugging tactics employed by participants; we describe some in detail as follows:

\topic{Retrieval} While most participants focused on improving retrieval first (P1-P6, P8-P12), some lacked clear intuitions on how to optimize it (P2, P6, P9, P11). Those with stronger intuitions typically increased chunk size (P1, P3-P5, P8, P10, P12), believing larger chunks provide better semantic coherence. However, P5 noted that without \sysname, such tuning would be financially impractical.

\topic{LLM Generation} Developers approached LLM debugging through ad-hoc exploration and iterative refinement, similar to behaviors observed in prior work~\cite{arawjo2023chainforge, zamfirescu2023johnny}. All participants quickly inspected prompt changes, though none used concrete evaluation metrics. Several (P5, P8, P11) expressed interest in more ``systematic'' approaches to prompt engineering, though their definitions of ``systematic'' varied considerably. For example, P8 wanted to review prompt diffs side by side in a spreadsheet for some set of queries, for some set of prompt perturbations. P5 and P11 simply wanted larger test sets of queries, perhaps to feel more systematic ``vibes.''

Component interdependence became evident when participants addressed Stage 3 of the task, which required supporting erroneous and ambiguous queries. Many had  to implement query rewriting to clarify and fix any typos in the query, using an additional LLM step (P4, P5, P7, P8, P12). Similarly, when increasing chunk size, developers had to consider implications for the LLM's context window limit (P4, P5). Participants who failed to account for these dependencies encountered issues like context overflow, requiring architectural reconsideration.

\section{Discussion}
\label{sec:discussion}

Here, we discuss the implications of our work. We first reflect on how to design RAG development tools, then situate our work within the landscape of debugging tools for agent-based systems, and finally acknowledge limitations.

\topic{Implications for Designing RAG Developer Tools} A common sentiment from participants was that \sysname enabled interactive changes they couldn’t previously make. This enthusiasm highlights a major gap in current RAG workflows: most vector databases lock developers into early decisions about chunking and embeddings, making iteration slow and costly. To better support experimentation, future systems could offer ``sandbox'' modes for lightweight, on-the-fly index creation. For instance, maybe vector databases could implement tiered storage approaches: maintaining pre-computed indexes for common configurations while enabling on-demand creation of experimental indexes. This would better support how developers naturally work---starting with intuition and refining designs through trial and error.

Moreover, visualization was central to effective iteration. Developers consistently examined retrieval quality before LLM outputs, and emphasized the importance of viewing both retrieved and non-retrieved chunks. Echoing prior work in data and traditional ML debugging~\cite{lee2020towards, hohman2018visual}, our findings suggest that RAG tools should support multiple, comparative views of retrieved content.  For example, while distribution views showing similarity scores helped identify poor chunk differentiation, participants also wanted to compare retrieval strategies (e.g., embedding-based vs. keyword-based) side by side---suggesting value in split-screen or overlay interfaces. While traditional ML debugging tools have supported visualizing and making sense of provenance for features, or numerical, structured data~\cite{bauerle2022symphony, wexler2019if}, RAG pipelines require tracing retrieved text back to its full unstructured {\em document} context. Embedding a document viewer within the interface, showing retrieved chunks in context with surrounding text, would help developers assess whether retrieved content preserves the right information or omits critical details. 

While participants appreciated \sysname's support for saving answers, they wanted more robust evaluation tools. This mirrors findings in data science and ML, where experiment tracking infrastructure is key to iteration~\cite{wang2019human, zaharia2018accelerating, vartak2016modeldb}. However, RAG development introduces new challenges: unlike traditional ML, there are no clear numerical metrics like accuracy or loss. While \sysname helped participants quickly scan outputs to assess correctness, manual review has clear limitations when pipelines scale. Automating correctness checks is difficult: an answer might be factually accurate but incomplete, or comprehensive yet include irrelevant details. Future tools must support evaluation frameworks that reflect these nuanced, domain-specific definitions of ``good'' quality.

\topic{RAG Debugging in the Landscape of Agent Tools} RAG pipelines increasingly operate within broader agent-based architectures, where they serve as only one component in workflows that may include document creation, database queries, and other tool-based interactions~\cite{epperson2025interactive, dibia2024autogen}. Building agentic workflows remains highly bespoke, with interfaces needing to adapt to the unique set of tools, tasks, and data that each pipeline processes. While \sysname provides valuable primitives for RAG-specific debugging, the question remains whether these primitives can extend to more complex agent architectures, and how to expand the set of primitives. \citet{epperson2025interactive} addresses some of these challenges by enabling interactive probing of multi-agent teams, allowing developers to modify messages exchanged between agents and explore counterfactual scenarios. But even still, developers struggle to quickly evaluate whether their edits result in correct outcomes, especially across multiple queries simultaneously like we observed with \sysname. Perhaps ``LLM call'' or ``message'' are too coarse-grained of primitives, and moreover, perhaps we should consider more sophisticated debugging approaches beyond simple ``edit-and-observe'' patterns. Future debugging interfaces could combine direct manipulation with ``intelligent'' sensemaking assistance~\cite{suh2023sensecape, gero2024supporting}---e.g., use LLMs to synthesize domain-specific debugging primitives based on system architecture and generate corresponding visualizations. For example, in a customer support chatbot, an LLM could analyze retrieved support documents to generate a network visualization showing related issues that customers typically encounter together, based on customer usage patterns. When the RAG pipeline provides incorrect or incomplete answers, developers could query this visualization to explore alternative documents; essentially, there are opportunities to build tools that help developers identify ``blind spots'' in their retrieval strategies.

\topic{Limitations and Future Work} While our study offers insights into RAG development workflows, we acknowledge several limitations.  First, the one-hour study duration does not reflect the full complexity of production RAG development, where developers often fine-tune embeddings or adopt more advanced techniques---better captured via longer-term contextual inquiry. Second, we focused on interactive debugging, not earlier stages like document ingestion (e.g., from PDF to chunks). Future work should explore tooling for preprocessing, cleaning, and embedding generation. Third, our participants varied in IR and ML expertise. While this reflects the broadening audience for RAG tools, some lacked familiarity with established frameworks, which may have limited how far they optimized their pipelines. Fourth, while participants made similar architectural changes (e.g., adding query rewriting), they often gave different or incorrect rationales. This suggests that our system did not fully support {\em interpretability}---while developers could observe the effects of changes, they sometimes developed incorrect intuitions about why certain modifications improved performance. Future tools could help developers form accurate mental models, potentially through causal explanations or grounded examples. Finally, our task design, though realistic in complexity, did not include latency or compute constraints common in production~\cite{shankar2024we}. These constraints often shape real-world decisions and may lead to different workflows. Future research should involve longer-term studies with production teams, and explore how debugging tools fit into ``full-stack'' RAG pipeline development and support users with diverse expertise.

\section{Conclusion}
\label{sec:conclusion}

RAG pipelines have become essential for building AI assistants with domain-specific knowledge, but developing effective pipelines remains challenging due to the deeply intertwined nature of retrieval and generation components. Our tool, \sysname, addresses these challenges by combining Python-based composable primitives with an interactive debugging environment that enables real-time parameter adjustments and immediate visualization of effects throughout the pipeline. Through our user study with 12 experienced practitioners, we observed how developers approach RAG pipeline development. Our findings highlight opportunities for future RAG development tools that better support rapid experimentation, systematic evaluation, and improved provenance tracking---ultimately enabling more robust AI assistants that can reliably access and reason with domain-specific knowledge.

\begin{acks}
We would like to thank the UC Berkeley EECS SUPERB REU program for supporting this project. We would also like to thank the EPIC Data Lab for their invaluable ideas and support throughout. 
\end{acks}

\bibliographystyle{ACM-Reference-Format}

\clearpage
\appendix
\section{Formative Interview Study Protocol} \label{app:formative-interview}

This appendix contains the semi-structured interview protocol used for our formative interviews with RAG practitioners. All interviews were conducted with approval from our institution's Institutional Review Board (IRB).

\subsection*{Introduction (2-3 minutes)}
\begin{itemize}
    \item Researcher introductions
    \item ``The purpose of this interview is to explore ways to improve interfaces for building RAG pipelines. Your responses in this interview will help us design better tools.''
    \item ``This is not an evaluation and we are here to learn from you.''
    \item Consent: ``Did you receive and sign the consent form?''
    \item Risks \& Benefits: ``There are no risks by participating in this study outside of everyday life. Participation is completely voluntary and you can stop the interview at any time.''
\end{itemize}

\vspace{1em}
\noindent Example script: \\
``Thank you for being a part of this study. We are researchers from [Institution] investigating ways to help developers building RAG pipelines do it better through visualizations or interface design. Your insights in this interview will help us design better tools. This interview will be from 45-60 minutes long. Your participation is completely voluntary and you may stop the interview at any time. There are no risks outside of everyday life by participating in this interview. We will be taking notes and making an audio recording for the duration of the interview. Do you consent to participating in this interview and being audio recorded?''

\subsection*{Background and Rapport Building (5 minutes)}
\begin{itemize}
    \item Could you tell me more about yourself and what you do?
    \item What kind of tasks are you building RAG pipelines for?
    \item Bridge question: What was the RAG system that you started with?
\end{itemize}

\subsection*{Topics (45-50 minutes)}

\subsubsection*{System Description}
\begin{itemize}
    \item (If the system is in production) Can you describe the RAG system you started with?
    \item How long did it take to get that to production?
    \item What was the biggest challenge getting the system production ready?
\end{itemize}

\subsubsection*{Architecture/Parameter Decisions \& Experimentation (25-30 minutes)}
\begin{itemize}
    \item What are the parameters of your RAG system?
    \item How many documents are you retrieving from?
    \item (Indexing): How do you come up with chunk size?
    \begin{itemize}
        \item What methods are you using to optimize chunk size?
        \item Have you experimented with different indexing methods?
    \end{itemize}
    \item (Retrieval): What retrieval methods are you using (e.g. Re-Rank, RankGPT, RAG-Fusion, Corrective-RAG)? How are you deciding to use that retrieval method?
    \begin{itemize}
        \item How do you experiment with different retrieval methods?
        \item How do you decide how many documents to retrieve?
    \end{itemize}
    \item (Generating) Once you have retrieved context, how are you prompting?
    \begin{itemize}
        \item How do you decide to construct prompt templates?
    \end{itemize}
    \item Can you walk me through the typical architecture of a pipeline you use for integrating RAG techniques into your LLM application?
    \begin{itemize}
        \item Was this the architecture that you started the system with?
        \item How/When are you deciding to evolve your architecture?
    \end{itemize}
    \item Tell me about a recent challenge you faced when integrating RAG with other systems or technologies in your LLM pipeline? Collaborating with other engineers? Integrating into larger systems?
    \begin{itemize}
        \item Follow-up: How did you approach that challenge?
    \end{itemize}
\end{itemize}

\section{Task Coding Cheat Sheet}
\label{app:task-cheat-sheet}

\noindent\fbox{%
    \parbox{\linewidth}{%
        \textbf{Task Coding Cheat Sheet:} This cheat sheet was provided to every participant during the interview as a markdown file for easy navigation. It has been reformatted for readability.
    }%
}

\subsection*{Quick Links}
\begin{itemize}
    \item \textbf{\hyperref[sec:run-the-pipeline]{Run the pipeline}} \vspace{0.3em}
    \item \textbf{\hyperref[sec:retriever]{Retriever}} \vspace{0.3em}
    \item \textbf{\hyperref[sec:llm]{LLM}} \vspace{0.3em}
    \item \textbf{\hyperref[sec:query]{Query}} \vspace{0.3em}
    \item \textbf{\hyperref[sec:answer]{Answer}} \vspace{0.3em}
    \item \textbf{Prompt Templates and Patterns:} 
    \begin{itemize}
        \item \hyperref[sec:formatting]{Formatting} \vspace{0.3em}
        \item \hyperref[sec:parse-json]{Parse JSON Output} \vspace{0.3em}
        \item \hyperref[sec:task-decomposition]{Task Decomposition} \vspace{0.3em}
        \item \hyperref[sec:re-rank]{Re-rank}
    \end{itemize}
\end{itemize}

\subsection*{Run the Pipeline}
\label{sec:run-the-pipeline}
\begin{lstlisting}[style=python]
python task/pipeline.py
\end{lstlisting}

\noindent\textbf{Retriever}
\label{sec:retriever}
\begin{lstlisting}[style=python]
retriever.invoke(
    query="Your query here",
    k=10,
    chunkSize=1000,
    chunkOverlap=10,
    retrievalMode="vanilla",
)
\end{lstlisting}

\noindent\textbf{LLM}
\label{sec:llm}
\begin{lstlisting}[style=python]
llm(
    prompt="this is the prompt",
    max_tokens=200,
    temperature=0.20
)
\end{lstlisting}

\noindent\textbf{Query}
\label{sec:query}
\begin{lstlisting}[style=python]
Query("This is the query string")
\end{lstlisting}

\noindent\textbf{Answer}
\label{sec:answer}
\begin{lstlisting}[style=python]
Answer("This is the answer string")
\end{lstlisting}

\vspace{1em}
\noindent\rule{\linewidth}{0.5pt}  %
\vspace{0.5em}

\subsection*{Prompt Templates and Patterns}  
These prompts serve as building blocks for common RAG pipeline formats. Adjust them as needed.

\noindent{\textbf{Formatting}}  
\label{sec:formatting}
\begin{lstlisting}[style=python]
FORMAT="""
You are an expert at answering questions given relevant context.
Given a question and context, answer the question according to the context.
Question: {question}
Context: {context}
"""
prompt = FORMAT.format(question="Example question", context="Example context")
\end{lstlisting}

\noindent{\textbf{Parse JSON Output}}  
\label{sec:parse-json}
\begin{lstlisting}[style=python]
FORMAT="""
Answer this question: {question}

Follow this JSON output:
{{"answer": []}}
"""
answer = llm(prompt=FORMAT.format(question=question), max_tokens=200, temperature=0.2)
json_decomp = json.loads(answer)
parsed_answer = json_decomp["answer"]
\end{lstlisting}

\noindent{\textbf{Task Decomposition}}  
\label{sec:task-decomposition}

\begin{lstlisting}[style=python]
QUERY_DECOMP="""
You are an expert at converting user questions into sub-questions.
Perform query decomposition.

Follow this JSON output:
{{"sub_questions": []}}

Question: {inputQuery}
"""
decomp = llm(prompt=QUERY_DECOMP.format(inputQuery=inputQuery), max_tokens=200, temperature=0.2)
json_decomp = json.loads(decomp)
sub_questions = json_decomp["sub_questions"]
\end{lstlisting}

\noindent{\textbf{Re-rank}}  
\label{sec:re-rank}
\begin{lstlisting}[style=python]
RERANK="""
You are an expert at ranking relevant context.

Question: {query}
Context: {context}
"""
docs_and_scores = [("context_1", 0.12), ("context_2", .34), ("context_3", 0.56)]
context = "\n".join([doc for doc, _ in docs_and_scores])
rerank = llm(prompt=RERANK.format(query=question, context=context), max_tokens=100, temperature=0.1)
\end{lstlisting}

\section{User Study Protocol for Raggy}
\label{app:raggy-user-study}

This appendix contains the semi-structured interview protocol used for our user study on \sysname. 
The study consists of contextual inquiry and structured interviews to observe how participants interact with \sysname and iterate on RAG pipelines. 
All interviews were conducted with approval from our institution's Institutional Review Board (IRB).

\subsection*{Phase 1: Onboarding}
\begin{itemize}
    \item The participant shares their screen in \textbf{VSCode}.
    \item The researcher introduces \sysname and explains the study’s purpose.
\end{itemize}

\subsection*{Phase 2: Task}
\begin{itemize}
    \item Participants complete the assigned task while sharing their screen in \textbf{VSCode}.
    \item The researcher observes participants’ workflow, taking note of interactions, challenges, and decision-making processes.
    \item Participants are encouraged to think aloud while using \sysname.
\end{itemize}

\subsection*{Phase 3: Interview Questions}

\subsubsection*{Probing UX Comments}
\begin{itemize}
    \item What aspects of the system are you \textbf{willing to wait for}, and what aspects are you \textbf{not willing to wait for}?
    \item What level of \textbf{quality} do you find \textbf{acceptable}, and is considered \textbf{unacceptable} when evaluating pipeline responses?
    \item What was the worst part of using \sysname?
    \item What was helpful about \sysname?
\end{itemize}

\subsubsection*{Workflow Evaluation}
\begin{itemize}
    \item \sysname prescribes a \textbf{specific workflow} for iterating on RAG pipelines.
    \begin{itemize}
        \item \textbf{What do you like} about this workflow?
        \item \textbf{What don’t you like} about this workflow?
    \end{itemize}
    \item How does this workflow \textbf{compare} to how you currently debug RAG pipelines?
    \item What aspects of this workflow were \textbf{helpful}? What aspects \textbf{weren’t}?
\end{itemize}

\subsection*{Closing Remarks}
\begin{itemize}
    \item Encourage participants to share any additional thoughts or suggestions.
    \item Thank the participant for their time and feedback.
\end{itemize}

\end{document}